\documentclass[floatfix,aps,prl,twocolumn,superscriptaddress,showpacs]{revtex4-1}
\usepackage{float}
\usepackage{color}
\usepackage{graphicx}
\usepackage{ulem}
\usepackage{epstopdf}
\usepackage{amsfonts}
\usepackage{amsmath}
\usepackage{mathtools}
\usepackage{bbold}
\usepackage[toc,page]{appendix}

\newcommand{\czus}[2]{\genfrac(){0pt}{0}{#1}{#2}}

\newcommand{\ec}{\overline{c}}
\newcommand{\eC}{\overline{\mathbf{C}}}
\newcommand{\eCr}{\overline{\mathbf{C}}_{[{\rm r},{\rm r}]}}
\newcommand{\nc}{c^{*}}
\newcommand{\nC}{\mathbf{C}^{*}}
\newcommand{\mean}[1]{\left\langle #1 \right\rangle}

\newcommand{\x}{\mathbf{x}}
\newcommand{\xr}{\mathbf{x}_{\rm r}}
\newcommand{\xl}{\mathbf{x}_{\rm l}}
\newcommand{\A}{\mathbf{A}}
\newcommand{\Ar}{\mathbf{A}_{[{\rm r},{\rm r}]}}
\newcommand{\Aeff}{\mathbf{A}_\textup{eff}}
\newcommand{\B}{\mathbf{B}}
\newcommand{\C}{\mathbf{C}}
\newcommand{\Cr}{\mathbf{C}_{[{\rm r},{\rm r}]}}
\newcommand{\D}{\mathbf{D}}
\newcommand{\Dr}{\mathbf{D}_{[{\rm r},{\rm r}]}}

\newcommand{\Fr}{\mathbf{f}_{\rm r}}
\newcommand{\I}{\mathbf{I}}
\newcommand{\J}{\mathbf{j}}
\newcommand{\Jr}{\mathbf{j}_{\rm r}}
\newcommand{\Om}{\mathbf{\Omega}}
\newcommand{\Omr}{\mathbf{\Omega}_{\rm r}}

\newcommand{\pdx}{\widetilde{\partial}_{1}^2}
\newcommand{\pdy}{\widetilde{\partial}_{2}^2}

\newcommand\chout{\bgroup\markoverwith{\textcolor{red}{\rule[0.5ex]{2pt}{1.0pt}}}\ULon}
\newcommand\fedout{\bgroup\markoverwith{\textcolor{blue}{\rule[0.5ex]{2pt}{1.0pt}}}\ULon}
\newcommand\grzout{\bgroup\markoverwith{\textcolor{green}{\rule[0.5ex]{2pt}{1.0pt}}}\ULon}

\DeclareMathOperator{\Tr}{Tr}
 
\begin{document}

\title{Non-equilibrium scaling behaviour in driven soft biological assemblies}

\author{Federica Mura}
    \thanks{These authors contributed equally}
\affiliation{Arnold-Sommerfeld-Center for Theoretical Physics and Center for
  NanoScience, Ludwig-Maximilians-Universit\"at M\"unchen,
   D-80333 M\"unchen, Germany.}
\author{Grzegorz Gradziuk}
    \thanks{These authors contributed equally}
\affiliation{Arnold-Sommerfeld-Center for Theoretical Physics and Center for
  NanoScience, Ludwig-Maximilians-Universit\"at M\"unchen,
   D-80333 M\"unchen, Germany.}
\author{Chase P. Broedersz}
\email{C.broedersz@lmu.de}
\affiliation{Arnold-Sommerfeld-Center for Theoretical Physics and Center for
  NanoScience, Ludwig-Maximilians-Universit\"at M\"unchen,
   D-80333 M\"unchen, Germany.}

\date{\today}

\begin{abstract}
Measuring and quantifying non-equilibrium dynamics in active biological systems is a major challenge, because of their intrinsic stochastic nature and the limited number of variables accessible in any real experiment. We investigate what non-equilibrium information can be extracted from non-invasive measurements using a  stochastic model of soft elastic networks with a heterogeneous distribution of activities, representing enzymatic force generation. In particular, we use this model to study how the non-equilibrium activity, detected by tracking two probes in the network, scales as a function of the distance between the probes. We quantify the non-equilibrium dynamics through the cycling frequencies, a simple measure of circulating currents in the phase space of the probes. We find that these cycling frequencies exhibit power-law scaling behavior with the distance between probes. In addition, we show that this scaling behavior governs the entropy production rate that can be recovered from the two traced probes. Our results provide insight in to how internal enzymatic driving generates non-equilibrium dynamics on different scales in soft biological assemblies.

\end{abstract}
\maketitle
\noindent 
Cells and tissue constitute a class of non-equilibrium many-body systems~\cite{Gnesottoreview,Fodorreview,Needleman2017,Schmidt2010,Jülicherreview}. Indeed, non-equilibrium activity has been observed in various biological systems, including membranes~\cite{Betz2009,Turlier2016}, chromosomes~\cite{Weber}, and the cytoplasm~\cite{Lau2003,Guo2014,Fakhri2014}. A distinguishing physical feature of such biological assemblies is that they are driven out of equilibrium collectively by internal enzymatic processes that break detailed balance at the molecular scale. The active nature of living matter on larger scales can be determined non-invasively by observing the steady-state stochastic dynamics of mescoscopic degrees of freedom using time-lapse microscopy experiments: The non-equilibrium dynamics of these systems can manifest as circulating probability currents in a phase space of mesoscopic coordinates~\cite{Battle2016, Gladrow2016,Gladrow2017,Gnesottoreview}. However, it remains unclear how such non-equilibrium measures  depend on the spatial scale on which the measurement is performed. This issue is not only of practical relevance in an experiment, it is also  of fundamental importance: a theoretical understanding of the spatial scaling behavior of broken detailed balance in internally driven systems may reveal how to extract quantitative information from  measurable phase space currents about the active nature of the system.

\begin{figure}[h!]
\centering
  \includegraphics[width=7cm]{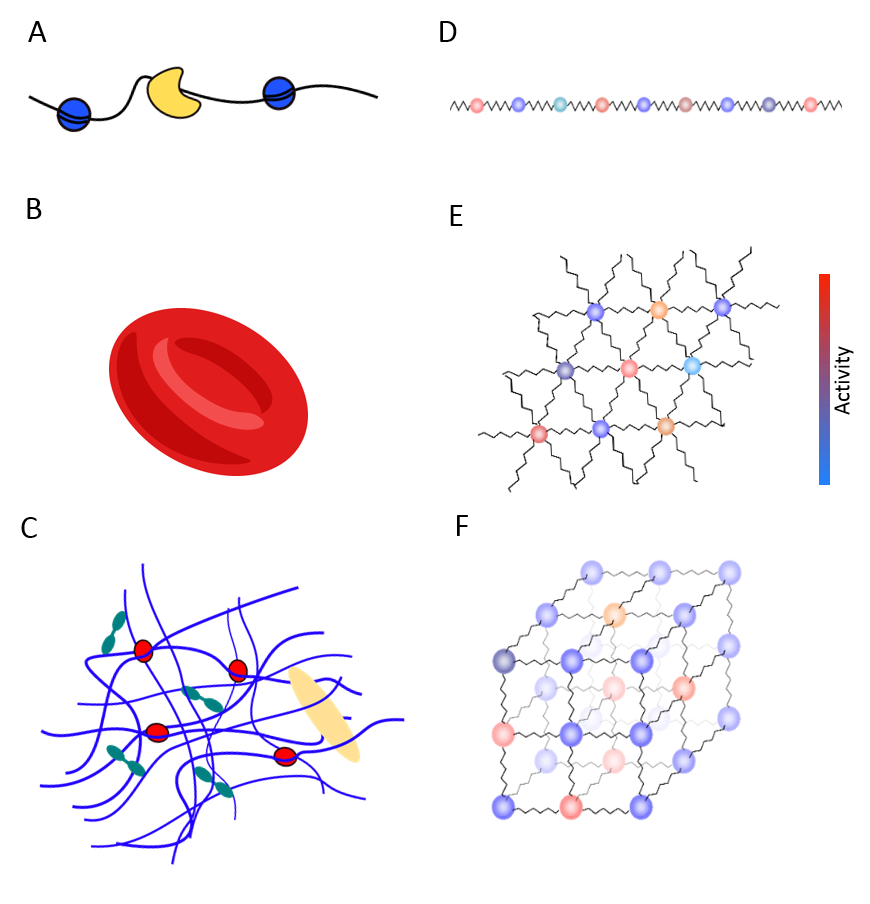}
  \caption{Schematic illustrating soft viscoelastic networks with heterogeneous driving for various types of cellular systems. A) chromosome B) red blood cell membrane C) cytoskeletal network with in D-F associated bead-spring models with heterogeneous active driving. The color of the bead indicates the intensity of activity, representing the variance (increasing from blue to red) of the associated active noise process. 
}
    \label{fig:bead_FDT}
\end{figure}
Here we consider a simple, yet general model for an internally driven elastic assembly to study non-equilibrium scaling behavior. This assembly is driven out of equilibrium by heterogeneously distributed stochastic forces, representing internal enzymatic activity (Fig.~\ref{fig:bead_FDT}). We quantify the non-equilibrium dynamics of such an assembly by the cycling frequencies associated to steady-state circulating currents in phase space~\cite{Gladrow2016,Gladrow2017}. To study how broken detailed balance manifests on different scales in a given system, we investigate how the cycling frequency of a pair of tracer probes depends on the spatial distance between these probes. Interestingly,  the cycling frequencies in our model exhibit a power-law scaling with the distance between probes with an exponent that depends on the dimensionality of the system. To provide a conceptual understanding of this scaling behavior, we develop an analytical calculation of these exponents. Furthermore, we show that the exponent associated to the power law of the cycling frequencies also underlies the scaling behavior of the  entropy production rate that can be recovered from measured trajectories. Therefore, we provide a framework to study the spatial scaling behavior of non-equilibrium measures in  soft elastic assemblies. 

Our model consists of a $d$-dimensional elastic network of $N$ beads, immersed in a simple Newtonian liquid at temperature $T$~\cite{Yucht2013,Broedersz2014,Mao2018}. We assume a lattice structure where each bead is connected to its nearest neighbours by springs of elastic constant $k$, as illustrated in Fig.~\ref{fig:bead_FDT}. For simplicity, we model  internal enzymatic activity by a Gaussian white noise with variance $\alpha _i$ at bead $i$. By assuming white noise, we effectively consider the dynamics of biological systems on time scales much longer than the characteristic timescales of the active processes~\cite{Gladrow2016, MacKintosh2008, Ruostekoski2003}. Importantly, these activity amplitudes, $\alpha _i\geq0$, are spatially heterogeneous, reflecting a spatial distribution of active processes in the system. These activity amplitudes are drawn independently from a distribution $p_\alpha$ with mean $\bar{\alpha}<\infty$ and standard deviation $\sigma_\alpha<\infty$ for each realization of the system. This description of a heterogeneously driven assembly is similar to bead-spring models in which the beads are coupled to distinct heat baths at different temperatures~\cite{Falasco2015,Bonetto2004,Rieder1967}. 

The temporal evolution of the probability distribution, $p(\x,t)$, of the beads' displacements $\x$, relative to their rest positions, is governed by a Fokker-Planck equation:
\begin{align}
\begin{split}
\frac{\partial p(\x,t)}{\partial t} &= - \nabla \cdot [\A\x p(\x,t)]+\nabla \cdot \D\nabla p(\x,t),\\
&=- \nabla \cdot  \J(\x,t) 
\label{eq:Fokker}
\end{split}
\end{align}
where $\J(\x,t) =\A\x p(\x,t) + \D\nabla p(\x,t) $ is the probability current. Here, $\A$ is the elastic interaction matrix, incorporating all nearest neighbor spring interactions between beads; the mobility matrix is assumed to be diagonal to exclude hydrodynamic interactions between the beads, and is absorbed in $\A$. The diffusion matrix, $\D$, is diagonal with elements $d_{ij}= \delta _{ij}\frac{k_B(T+ \alpha _i)}{\gamma}$, where $\gamma$ is the damping coefficient describing the viscous interaction between a bead and the immersing liquid. 
The steady-state dynamics of this active network is described by 
\begin{equation}
p(\x) = \frac{1}{\sqrt{(2\pi)^{dN} \det \C}} e^{-\frac{1}{2}\x^T\C^{-1}\x},
\end{equation}
 where $\C= \langle \x \otimes \x \rangle $ is the covariance matrix, which can be obtained by solving the Lyapunov equation $\A\C~+~\C\A^T~ =~ -2\D$~\cite{Risken}. In the simplest limit, the activities are spatially homogeneous: $\alpha _i= \alpha $ $\forall \, i$, resulting in a dynamics that  reduce to that of an effective equilibrium scenario with $\J=0$. By contrast, in heterogeneously driven systems with non-identical $\alpha _i$'s, we obtain Non-Equilibrium Steady-State (NESS) dynamics with $\J \neq 0$~\cite{Risken}.

\begin{figure}[t]
\centering
  \includegraphics[width=8 cm]{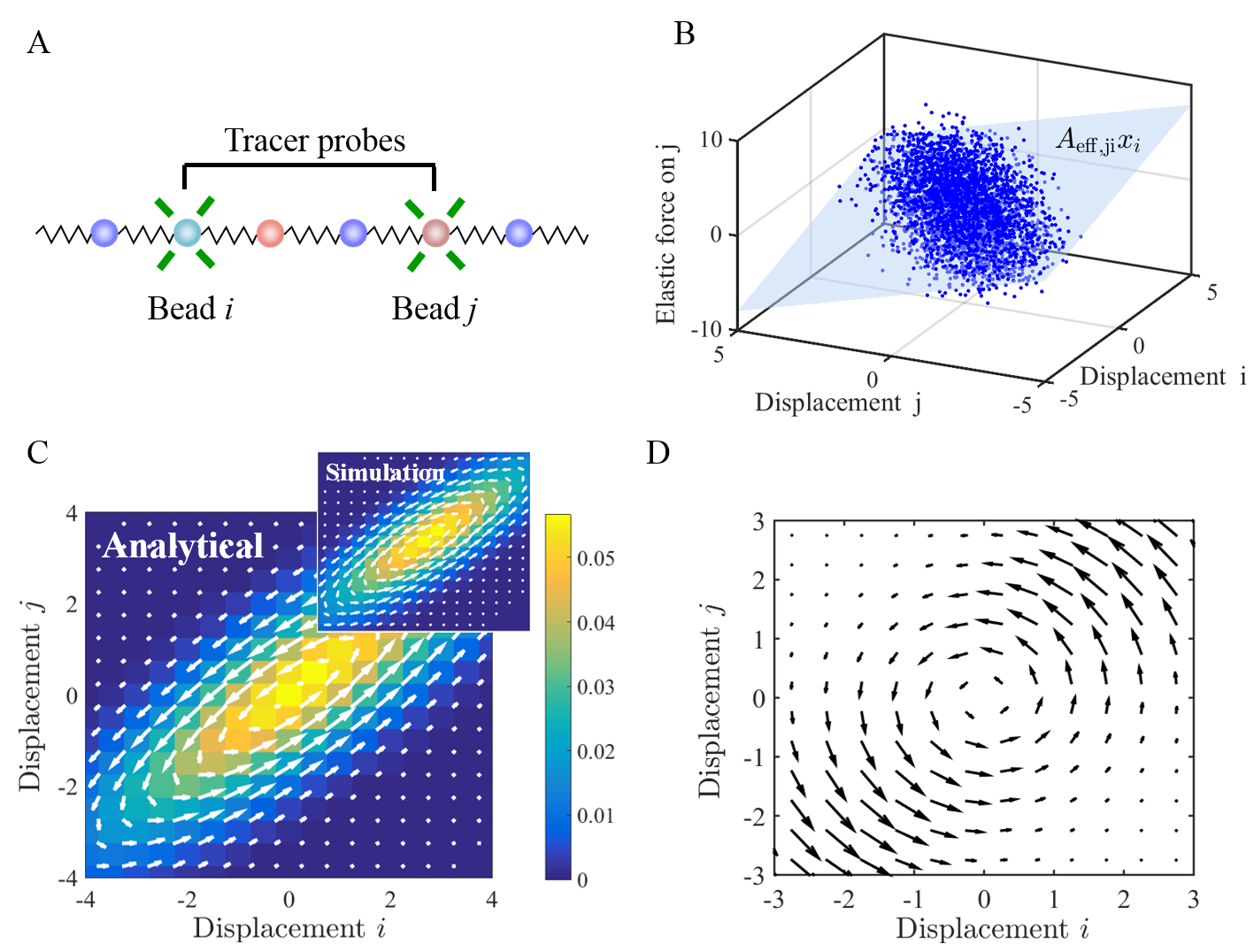}

  \caption{Reduced system of tracked probed. A) Schematic of two fluorescently labelled probe beads in a larger system. B) Elastic force acting on bead $j$ obtained at different time steps of a simulation of the Langevin dynamics of the  full system (blue points), and the effective linear force, $\Aeff\xr$,  from analytical calculations (light blue plane). C) Probability density (color map) and probability current (white arrows)  calculated analytically from the effective 2D system, together with results from simulating the full system in the inset.  D)  The non-conservative part of the effective force field: $\frac{(\Aeff - \Aeff^T)}{2}\xr$ (black arrows) can  contribute to the rotation in phase space in non-equilibrium systems.}

    \label{fig:reduceddfig}
\end{figure}

If we were able to observe the stochastic motion of all beads in the network, we could directly measure the full probability current $\J(\x)$ and extract information about the complete non-equilibrium dynamics of the system. However, in an actual experiment typically only a small subset of the degrees of freedom can be tracked (Fig.~\ref{fig:reduceddfig}A). What information on the non-equilibrium dynamics of the system can be extracted from such limited observations? To address this question, we investigate a scenario where only a few degrees of freedom are accessible. 

We start by reducing our description to the marginal distribution, $p_{\rm r}(\xr)~=~\int dx_{k\not\in [r]} p(x_1,x_2,..,x_{dN})$, of a subset $[\rm r]$ of $n$ tracked degrees of freedom $\xr$. By integrating out the subset $[{\rm l}]$ of $m$ unobserved degrees of freedom  $\xl$ on both sides of Eq.~\eqref{eq:Fokker} in the steady-state limit, we obtain (see supplementary material): 
\begin{equation}
0= - \nabla \cdot[\Aeff\xr p_{\rm r}(\xr)]+\nabla \cdot \Dr\nabla p_{\rm r} (\xr),
\label{eq:effectivefokker}
\end{equation}
where the sub-index $[{\rm r},{\rm r}]$ of a matrix indicates the sub-matrix corresponding to the reduced set of observed variables.
In addition, we introduce the effective linear interaction (Fig.~\ref{fig:reduceddfig}B), which can be written as $\Aeff\xr$, with $\Aeff= \Ar+ \A_{[{\rm r},{\rm l}]} \C_{[{\rm l},{\rm r}]}\Cr^{-1} $. Here,
$\A_{[{\rm r},{\rm l}]}$  and $\C_{[{\rm l},{\rm r}]}$ are rectangular matrices of sizes $[n \times m]$  and $[m \times n]$, given by the elements of indices $\rm [r,l]$ of $\A$ and $\rm [l,r]$ of $\C$, respectively.
Thus, we obtain an effective stationary Fokker-Planck equation for the reduced system (Eq.~\eqref{eq:effectivefokker}). By solving this equation, we obtain the exact steady-state reduced probability distribution $p_{\rm r}(\xr)$ and probability current density:
\begin{equation}
 \Jr (\xr)=\Aeff\xr p_{\rm r}(\xr) +\Dr \Cr^{-1}\xr p_{\rm r}(\xr), 
 \label{eq:effectivecurrent}
\end{equation}
which can, in principle, be measured  directly  from the trajectories of the observed degrees of freedom at steady state (Fig.~\ref{fig:reduceddfig}C). 

We can use this reduced description to investigate how broken detailed balance manifests at different scales in the network. In particular, we consider the simplest  case of a reduced system of only two tracked beads in a larger system, as illustrated in Fig.~\ref{fig:reduceddfig}A. It is convenient to quantify the probability currents in the 2D phase space of these two tracer beads by a pseudoscalar quantity: the average cycling frequency around the origin~\cite{Weiss2003,Gladrow2016,Gladrow2017}. For linear systems, we can express the reduced probability current as $\Jr(\xr)=\Omr \xr p_{\rm r}(\xr)$, where $\Omr$ is a 2D matrix with purely imaginary eigenvalues $\lambda= \pm i \omega$, with $\omega$ representing the cycling frequency. 

\begin{figure}
\centering
  \includegraphics[width=8cm]{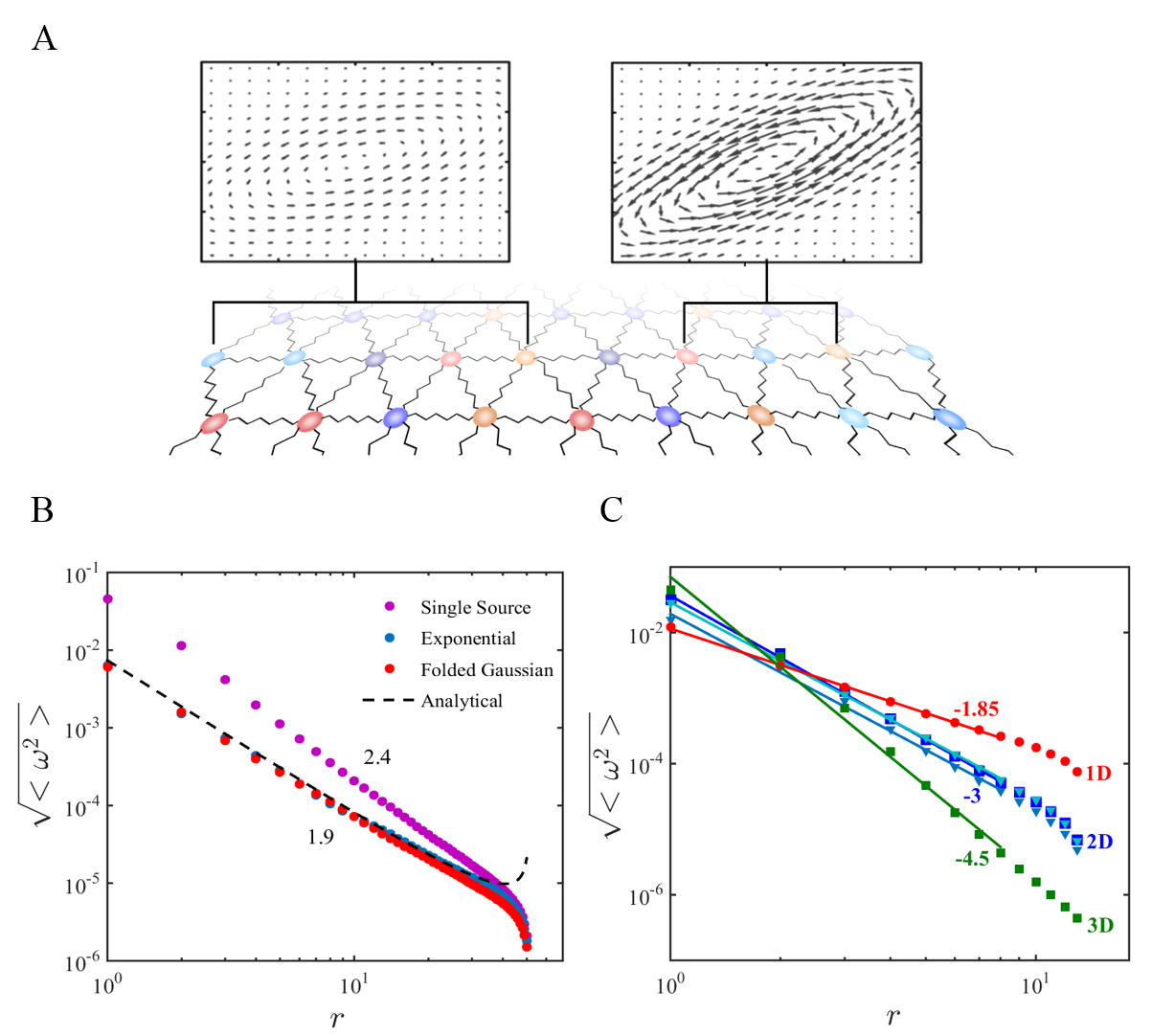}
  \caption{Spatial scaling behavior of cycling frequencies. A) Steady-state current cycles in phase space of two tracer beads for a nearby pair of probes (left) and distant pair of probes (right).  B) Scaling behavior of the cycling frequencies, $\sqrt{\langle\omega^2(r)\rangle}$, of a pair of probes beads as a function of their spatial distances, obtained for a 1D chain and different activity distributions, as indicated in the legend. C) Scaling behavior of the cycling frequencies, $\sqrt{\langle\omega^2(r)\rangle}$, obtained for different lattices and a folded Gaussian activity distribution. Triangular and square markers represent  triangular and square/cubic lattices, respectively.  Light/dark blue triangles represent triangular networks with zero/finite rest length springs. }    \label{fig:freqfig}
\end{figure}

This cycling frequency can be measured for a pair of probe beads at a distance $r$, and this frequency will depend on the specific configuration of the activity amplitudes $\alpha_i$ at all beads in the system. We aim to compute how this cycling frequency depends on $r$ after averaging over all activity configurations. Since $\omega$ is expected to be distributed symmetrically around $0$, we calculate $\sqrt{\langle\omega^2(r)\rangle_\alpha}$ for pairs of beads separated by a distance $r$. Here, the average $\langle...\rangle_\alpha$ is taken over an ensemble of activities $\{\alpha_i\}$ drawn from the distribution $p_\alpha$. Intuitively, the magnitude of the circulation of currents in phase space typically decreases with the distance between the probes, as shown in Fig.~\ref{fig:freqfig}A. This reduction of the circulation is reflected by a decrease of the cycling frequency $\omega$ with distance. Remarkably, $\sqrt{\langle\omega^2(r)\rangle_\alpha}$ appears to depend on the distance between the tracer beads, $r$, as a power law, $\sqrt{\langle\omega^2(r)\rangle_\alpha}\propto r^{-\mu}$, with $\mu\approx 1.9$ for a 1D chain with a folded Gaussian or an exponential distribution of activities, as depicted in Fig.~\ref{fig:freqfig}B. 

To investigate how the architecture of the system affects the scaling behavior of the cycling frequencies, we considered different network structures, including square, triangular, and  cubic lattices. In particular, we calculated $\sqrt{\langle\omega^2(r)\rangle_\alpha}$, where we also averaged over  different lattice directions.
Interestingly, we find that the characteristic exponent $\mu$ appears to depend strongly on the dimensionality of the lattice, but not on its geometry, as shown in Fig.~\ref{fig:freqfig}B-C. These results suggest that the distance dependence of the cycling frequency is determined in part by the long wavelength elastic properties of the system. Importantly, however, the scaling of cycling frequency is sensitive to the spatial structure of the activities. For example, in the simple case of a delta-distributed (single-source) activity on a 1D chain, we find $\mu_{\rm single}\approx 2.4$ (Fig.~\ref{fig:freqfig}B) in contrast to the value $1.9$ obtained above for spatially distributed activities. 

To obtain more insight into the scaling behavior of the cycling frequencies, we derive an analytical expression for the cycling frequency as a function of the distance between the observed beads, $\omega (r)$. In general, it can be shown that for a linear system described by a Fokker-Planck equation, the cycling frequencies are given by (see supplementary materials):
\begin{equation}
\omega_{ij}=\frac{1}{2\gamma}\frac{\langle \tau_{ij} \rangle}{\sqrt{\det\Cr}}
\label{eq:torque}
\end{equation}
 where $\tau_{ij}\coloneqq \x\times\Fr(\x)= x_if_j(\x)-x_jf_i(\x)$ is a generalized phase space torque in the $x_i$-$x_j$ plane, with $f_i(\x)$ denoting the deterministic force acting on the $i^{\rm th}$ bead. This result is intuitive: for an overdamped system the mean angular velocity is proportional to the mean torque and the factor $1/\sqrt{\det\Cr}$ ensures coordinate invariance.
For the 1D  chain of beads (Fig.~\ref{fig:bead_FDT}D), Eq.~\eqref{eq:torque} reduces to:
\begin{equation}
\omega_{ij}=\frac{k}{\gamma}\frac{\pdy c_{ij}}{\sqrt{\det\Cr}},
\label{eq:omchain}
\end{equation}
where $c_{ij}$ is the $i,j^{\rm th}$ element of the covariance matrix  $\C$, and with the discrete second derivative across rows denoted as: $\pdy c_{ij}=c_{i,j+1}-2c_{i,j}+c_{i,j-1}$. Thereby, we have reduced the problem of calculating $\omega(r)$ to finding the covariance matrix of the system.

The structure of $\D$ suggests a natural decomposition of the covariance matrix $\C$ into equilibrium ($\eC$) and non-equilibrium ($\nC$) contributions: $\C=(k_BT/k)\eC+(k_B\bar{\alpha}/k)\nC$, such that $\eC$ and $\nC$ are dimensionless. Both $\eC$ and $\nC$ can be found by solving the Lyapunov equation, which for the 1D chain is given by
\begin{align}
\pdx \ec_{ij}+\pdy \ec_{ij}&=-2\delta_{ij} \label{eq:diffec} \\
\pdx \nc_{ij}+\pdy \nc_{ij}&=-2\delta_{ij}\frac{\alpha_i}{\bar{\alpha}} \label{eq:diffnc},
\end{align}
where $\pdx$ indicates the discrete second derivative across columns. These equations represent discrete stationary diffusion equations, with sources of divergence given by  $\delta_{ij}$ and $\delta_{ij}(\alpha_i/\bar{\alpha})$, respectively. 
This result prescribes how a spatial distribution of activities structures the covariance matrix.

We can make further progress by noting that the principle of detailed balance imposes $\omega_{ij}=0$ at thermal equilibrium, which together with Eq.~\eqref{eq:omchain} implies $\pdy \ec_{ij}=0$. We can, therefore, substitute $\pdy c_{ij}$ in Eq.\eqref{eq:omchain} by $\pdy\nc_{ij}$, and then expand this equation up to linear order in $\bar{\alpha}/T$ to obtain
\begin{equation}
\omega_{ij}=\frac{k}{\gamma}\frac{\bar{\alpha}}{T}\frac{\pdy\nc_{ij}}{\sqrt{\det\eCr}}.
\label{eq:omegalin}
\end{equation}
We proceed by calculating $\nC$ for a given distribution of activities $\{\alpha_i\}$. Because of the linearity of Eq.~\eqref{eq:diffnc}, $\nC$ is a superposition of steady-state solutions to single-source problem, i.e. a delta-distribution for which all but one of the activities would be set to zero. Denoting  the element of $\nC$ at a distance $r$ from the single activity source by $\nc(r)$, we obtain the ``covariance current" $\partial_r \nc(r)\sim1/r$. Here we employed a continuous approximation of the discrete diffusion problem in Eqs.~\eqref{eq:diffec} and \eqref{eq:diffnc}. Thus, $\nc(r)= -a\ln(r)+b$ for a single-source problem with integration constants $a$ and $b$. Using this expression for $\nc(r)$ together with Eq.~\eqref{eq:omegalin}, we obtain for the single source case: $\omega^2_{\rm single}(r)=\frac{k^2}{\gamma^2}\frac{\alpha^2}{T^2}\frac{a^2}{r^4}\frac{1}{\det\eCr(r)}$, where $\alpha$ is the source's activity.

Next, we use a superposition of  single source solutions for $\nc(r)$ to obtain the non-equilibrium contribution of the covariance matrix $\nC$ for a specific configuration of many activity sources $\{\alpha_i\}$. Using this result in conjunction with Eq.~\eqref{eq:omegalin} and performing an ensemble average over the distribution of activity realizations, we arrive at the central result 
\begin{equation}
\langle\omega^2(r)\rangle_\alpha=\frac{k^2}{\gamma^2}\frac{ \sigma ^2 _{\alpha}}{T^2}\frac{\pi a^2}{2r^3}\frac{1}{\det\eCr(r)}.
\label{eq:scaling}
\end{equation}
Finally, we note that the elements of the equilibrium covariance matrix are given by $\ec _{i,j}=\textup{min}(i,j) -ij/(N+1)$, and find that for $r\ll N$, $\det\eCr(r)$ exhibits  a power law behavior, $\det\eCr(r)\sim r$. Therefore, from this analysis we find for a 1D chain with heterogenous activities  $\mu=2$, independent of the activity distribution $p_\alpha$. Furthermore, we find $\mu_{\rm single}=2.5$ for a single-source activity, in accord with our numerical result (see Fig.\ref{fig:freqfig}B).
 This calculation provides insight into how a combination of features of the equilibrium and non-equilibrium contributions to the covariance matrix determine the spatial scaling behavior of cycling frequencies.

Non-zero cycling frequencies directly reflect broken detailed balance, suggesting a connection between $\omega$ and measures of the internal driving, including the rate of entropy production. For a Markovian system described by a Fokker-Planck equation, the total entropy production rate under steady-state conditions is given by~\cite{Seifertreview}:
\begin{equation}
\Pi_\textup{tot}=k_{\rm B}\int d\x\frac{\J^T(\x)\D^{-1}\J(\x)}{p(\x)},
\label{eq:entropyFokker}
\end{equation}
where $k_{\rm B}$ is Boltzmann's constant. The validity of this result relies on the equivalence between the Fokker-Planck and Langevin descriptions. However, we have seen that the marginal probability density of the reduced system is only described by a Fokker-Planck equation (see Eq.~\eqref{eq:effectivefokker}) at steady state, reflecting the loss of Markovianity after coarse-graining. Nonetheless, we can define an effective dynamics of the reduced set of variables through the Langevin equation
\begin{equation}
\frac{d\xr (t)}{dt}=  \Aeff\xr (t) + \sqrt{2\Dr}\,\mathbf{\xi}_{\rm r}(t),
\label{eq:effectivelangevin}
\end{equation}
with  Gaussian white noise $\mathbf{\xi}_{\rm r}(t)$.
This equation of motion results in the exact steady-state probability and current densities, but with an approximate stochastic dynamics. In particular, the effective interaction matrix $\Aeff$ (see Eq.~\eqref{eq:effectivefokker}) captures only the average interaction between the traced variables, as illustrated in Fig.~\ref{fig:reduceddfig}B. Furthermore, in contrast to the full deterministic forces ($\A\x$), these effective interactions  (Fig.~\ref{fig:reduceddfig}C) need not to derive from a potential and, thus, may contain a non-conservative component (Fig.~\ref{fig:reduceddfig}D).  

The entropy production rate associated with the effective Markovian dynamics in Eq.(\ref{eq:effectivelangevin}) is given by
\begin{figure}[t]
\centering
  \includegraphics[width=8cm]{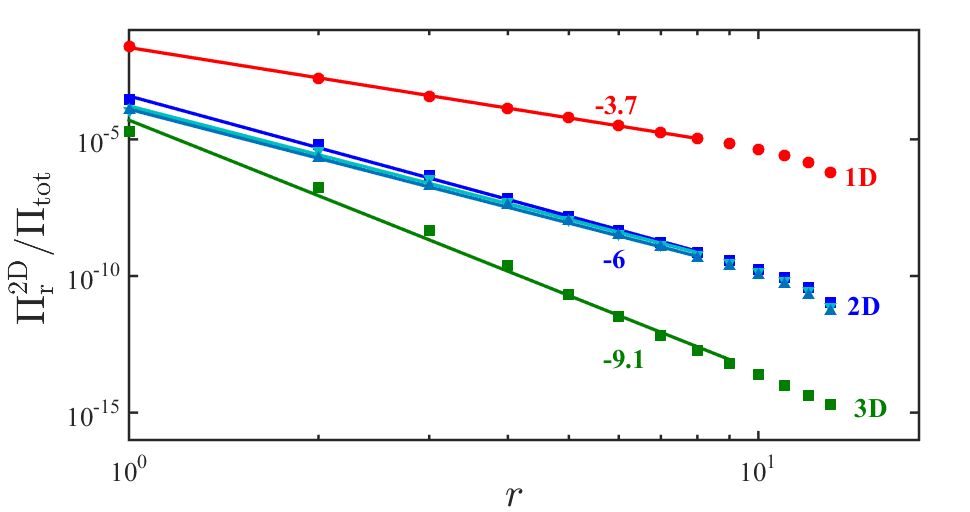}

  \caption{Spatial scaling behavior of the entropy production rate, $\Pi_{\rm r}^{\rm 2D}$, of a pair of probe beads as a function of their spatial distance $r$, obtained for different lattices and a folded Gaussian activity distribution. Note the entropy production rate of the reduced system is scaled by the total entropy production rate of the whole network, $\Pi_{\rm tot}$.  Triangular and square markers represent  triangular and square/cubic lattices, respectively.  Light/dark blue triangles represent triangular networks with zero/finite rest length springs.}
    \label{fig:entropyfig}
\end{figure}

\begin{equation}
\Pi_{{\rm r}}=k_B\int d\xr\frac{\Jr^T(\xr)\Dr^{-1}\Jr(\xr)}{p_{\rm r}(\xr)}\leq \Pi_\textup{tot},
\label{eq:entropyred}
\end{equation}
where $\Jr(\xr)$ is defined in Eq.~\eqref{eq:effectivecurrent}. Note, estimating $\Pi_{\rm r}$ by using the Markovian formalism allows us to set a lower bound for the total entropy production rate $\Pi_{\rm tot}$ (see supplementary materials), similar to what already shown  for discrete systems~\cite{Bisker2017}. In the 2D case with two traced degrees of freedom that we consider here, Eq.~\eqref{eq:entropyred} reduces to (see supplementary materials)
\begin{equation}
\Pi_{\rm r}^{\rm 2D}=k_B\omega ^2\Tr{(\Cr\Dr^{-1})}.
\label{eq:entropy2d}
\end{equation}
This result provides an explicit relation between the partial entropy production rate and the  cycling frequency $\omega$. Note, all quantities in the expression for $\Pi_{\rm r}^{\rm 2D}$ can be observed in an experiment, providing a direct way to non-invasively determine the reduced rate of entropy production for a set of traced degrees of freedom. Since $\Tr{(\Cr\Dr^{-1})}$ depends only weakly on $r$, as long as $1\ll r\ll N$, we expect a scaling behavior $\langle \Pi_{r}^{\rm 2D}\rangle \sim r^{-2 \mu}$. This result shows that the spatial scaling behavior of the cycling frequencies directly determines the spatial scaling behavior of the entropy production rate.

In summary, we here demonstrate theoretically how  experimental measures of non-equilibrium activity in internally driven linear networks are affected by the lengthscale at which the system is observed. Specifically, we developed a general framework to predict the scaling behavior of cycling frequencies and the entropy production rate that can be inferred by tracing pairs of degrees of freedom. We showed the exponent $\mu$ that governs this behavior for a system with heterogeneous random activities, is insensitive to the details of distribution of activities. However, this exponent depends sensitively on the dimensionality of the system. The predicted scaling behaviour can be tested in biological~\cite{Betz2009,Lau2003,Guo2014,Mizuno2007,Fakhri2014,Lieleg2010,Koenderink2009,Schaller2010,Jensen2015} and artifical~\cite{Palacci2013,Buttinoni2013} systems under non-equilibrium steady-state conditions.

\begin{acknowledgments}
We thank E.~Frey, J.~Gladrow, F.~Gnesotto, P.~Ronceray, and C.~Schmidt for many stimulating discussions.
This work was supported by the German Excellence Initiative via
 the program NanoSystems Initiative Munich (NIM), by a DFG Fellowship through the Graduate School of Quantitative Biosciences Munich (QBM). Part of this work was performed at the Aspen Center for Physics, which is supported by National Science Foundation grant PHY-1607611. \end{acknowledgments}

\onecolumngrid
\newpage
\begin{appendices}
\section{Derivation of Eq.~(\ref{eq:effectivefokker})}
\label{effectiveequation}
\noindent Here, we derive Eq.~\eqref{eq:effectivefokker}, which describes the steady state distribution of traced variables.
Integrating out the unobserved degrees of freedom on both sides of the Fokker-Plank equation (Eq.~\eqref{eq:Fokker}), and using the Einstein notation for summing over repeated indexes, we obtain:
\begin{equation}
\label{eq:FP3parts}
\overbracket{\int d\xl \partial_t p(\x)}^{(I)}=-\overbracket{\int d\xl \partial_i [a_{ij}x_j p(\x,t)]}^{(II)}+\overbracket{\int d\xl d_{ij}\partial_i\partial_j p(\x,t)}^{(III)}
\end{equation}
where $a_{ij}$ and $d_{ij}$ are the elements of the interaction matrix $\A$ and the diffusion matrix $\D$, respectively. Rewriting the probability as $p(\x,t)=p(\xl |\xr ,t)p_{\rm r}(\xr ,t)$, we can separately calculate each term in Eq.$(\ref{eq:FP3parts})$. The first term $(I)$ gives:
\begin{align}
\begin{split}
\int d\xl\partial_t p_{\rm r}(\xr ,t)p(\xl |\xr ,t)&=\partial_t p_{\rm r}(\xr,t)\int d\xl p(\xl |\xr ,t)=\partial_t p_{\rm r}(\xr ,t) 
\end{split}
\end{align}
For the second term (II), we obtain
\begin{align}
\begin{split}
\label{eq.partII}
\int d\xl \partial_i [p_{\rm r}(\xr ,t)p(\xl |\xr ,t)a_{ij}x_j]&=  \delta_{i,[{\rm r}]} \partial_i [p_{\rm r}(\xr,t)\int d\xl p(\xl |\xr ,t)a_{ij}x_j] \\
 &=\delta_{i,[{\rm r}]} \partial_i [ p_{\rm r}(\xr ,t)a_{ij} \mean{x_j|\xr ,t} ]
\end{split}
\end{align}
where $\delta_{i,[{\rm r}]}=1$ if $x_i$ is one of the observed coordinates and zero otherwise. In the first line we use that the probability density vanishes at infinity faster than $1/x$.  Similarly, the third term (III) can be written as
\begin{align}
\begin{split}
\label{eq.partIII}
\int d\xl d_{ij}\partial_i\partial_j [p_{\rm r}(\xr ,t)p(\xl |\xr ,t)] &=\delta_{i,[{\rm r}]}\delta_{j,[{\rm r}]} d_{ij}\partial_i\partial_j [p_{\rm r}(\xr ,t) \int d\xl p(\xl|\xr,t)]\\
&=\delta_{i,[{\rm r}]}\delta_{j,[{\rm r}]} d_{ij}\partial_i\partial_j p_{\rm r}(\xr ,t) 
\end{split}
\end{align}
An explicit calculation of the conditional averages appearing in Eq.(\ref{eq.partII}) yields $ \mean{\xl|\xr} = \C_{[{\rm l},{\rm r}]}\C_{[{\rm r},{\rm r}]}^{-1} \xr $~\cite{Tong}. We can substitute  contributions (I), (II) and (III) in Eq.~(\ref{eq:FP3parts}) under steady state conditions to obtain Eq.~(\ref{eq:effectivefokker}). 
\section{Derivation of Eq.~(\ref{eq:torque})}
\noindent Here we  derive the expression in Eq.~\eqref{eq:torque} for the cycling frequencies. To this end, we  first show that the right hand side of this equation is invariant under orientation preserving linear transformations restricted to the 2-dimensional reduced subspace. Let us consider such a transformation: $\xr'=\B\xr$, $\Fr'=\B\Fr$, and denote by $\Cr'$ the reduced covariance matrix in the transformed coordinates.
\begin{equation}
\label{eq:detB}
\B\Cr\B^T=\Cr'\quad\Longrightarrow\quad \det\B=\sqrt{\frac{\det\Cr'}{\det\Cr}}
\end{equation}
Using this result together with the transformation properties of the vector product, we obtain 
\begin{equation}
\frac{\mean{\tau_{ij}}}{\sqrt{\det\Cr}}=\frac{\mean{\xr\times\Fr(\x)}}{\sqrt{\det\Cr}}=\frac{\mean{\xr'\times\Fr'(\x')}}{\sqrt{\det\Cr}}\frac{1}{\det \B}=\frac{\mean{\tau'_{ij}}}{\sqrt{\det\Cr'}}.
\end{equation}
The coordinate invariance of this term allows us to specifically consider the convenient coordinates in which $\Cr=\I$:
\begin{equation}
\label{eq:taucalc}
\frac{1}{\gamma}\mean{\tau_{ij}}=\frac{1}{\gamma}\mean{\xr\times\Fr(\x) }=\frac{1}{\gamma}\int d\xr\mean{\xr\times\Fr(\x)|\xr}p_{\rm r}(\xr)=\frac{1}{\gamma}\int d\xr\ \xr\times\mean{\Fr(\x)|\xr}p_{\rm r}(\xr)
\end{equation}
We can further expand this expression by using $\Omr=\Aeff+\Dr\Cr^{-1}$. (The expression for $\Omr$ follows immediately from Eq.~\eqref{eq:effectivecurrent}, since we require $\Jr(\xr)=\Omr\xr p_{\rm r}(\xr)$.)
\begin{equation}
\label{eq:fcond}
\frac{1}{\gamma}\mean{\Fr(\x)|\xr}=\Aeff\xr=\Omr\xr-\Dr\Cr^{-1}\xr.
\end{equation}
Combining this result with Eq.~\eqref{eq:taucalc}, we arrive at
\begin{equation}
\label{eq:meantau}
\frac{1}{\gamma}\mean{\tau_{ij}}=\int d\xr\ \xr\times (\Omr\xr)p_{\rm r}(\xr)-\int d\xr\ \xr\times(\Dr\Cr^{-1}\xr)p_{\rm r}(\xr).
\end{equation}
Using the explicit form of $\Omr$ (see Eq.~\eqref{eq:omr}), we evaluate the first term in this expression,
\begin{equation}
\int d\xr\ \xr\times (\Omr\xr)p_{\rm r}(\xr)=\int d\xr\ \omega_{ij}(x_i^2+x_j^2)p_{\rm r}(\xr)=\omega_{ij}(c_{ii}+c_{jj})=2\omega_{ij}.
\end{equation}
In addition, we confirm by direct calculation, that, as expected, the second term in Eq.~\eqref{eq:meantau} vanishes:
\begin{align}
-\int d\xr\ \xr\times(\Dr\xr)p_{\rm r}(\xr)&=\int d\xr\ (-x_j,\ x_i)
\left(\begin{array}{cc}
d_{ii} & d_{ij} \\
d_{ij} & d_{jj} \\
\end{array}\right)\czus{x_i}{x_j}p_{\rm r}(\xr)=\\
&=\int d\xr\ [-d_{ii}x_ix_j-d_{ij}x_j^2+d_{ij}x_i^2+d_{jj}x_ix_j]p_{\rm r}(\xr)=\\
&=\underbrace{c_{ij}}_0 (d_{jj}-d_{ii})+d_{ij}\underbrace{(c_{ii}-c_{jj})}_0=0
\end{align}
Altogether, this gives us the desired result:
\begin{equation}
\frac{1}{2\gamma}\frac{\mean{\tau_{ij}}}{\sqrt{\det\Cr}}=\omega_{ij}
\end{equation}

\section{Derivation of Eq. (\ref{eq:entropyred})}
\noindent Here we show that $\Pi _\textup{tot} \ge \Pi_{\rm rr} $.
\begin{align}
\begin{split}
\frac{\Pi _\textup{tot}-\Pi_{\rm r}}{k_{\rm B}}
&=\int d\x\frac{\J^T(\x)\D^{-1}\J(\x)}{p(\x)}-\int d\xr\frac{\Jr^T (\xr)\Dr^{-1}\Jr (\xr)}{p(\xr)}\\ 
&=\frac{\gamma}{k_{\rm B}}\sum_{j\in [l]}\int d\x\frac{v_j^2(\x)}{(T+\alpha_j)}p(\x) + \frac{\gamma}{k_{\rm B}}\sum_{i\in [r]}\ \left[\left(\int d\x\frac{v_i^2(\x)}{(T+\alpha_i)}p(\x)\right)- \int d\xr \frac{\langle v_i(\x)|\xr \rangle^2}{(T+\alpha_i)}p(\xr)\right]   \\ 
&=\frac{\gamma}{k_{\rm B}}\left[\sum_{j\in [l]}\int d\x \frac{v_j^2(\x)}{(T+\alpha_j)}p(\x) +\sum_{i\in [r]}\ \int d\xr\left[\left(\int d\xl \frac{v_i^2(\x)}{(T+\alpha_i)}p(\xl|\xr)p(\xr)\right)-\frac{\langle v_i(\x)|\xr \rangle ^2}{(T+\alpha_i)} p(\xr) \right]\right]\\
&=\frac{\gamma}{k_{\rm B}}\left[\sum_{j\in [l]}\frac{\langle v_j^2(\x)\rangle}{(T+\alpha_j)}+ \sum_{i\in [r]}\ \int d\xr \underbrace{\left(\langle v_i^2(\x)|\xr\rangle - \langle v_i(\x)|\xr \rangle ^2 \right)}_{\geq 0} \frac{p(\xr)}{(T+\alpha_i)}\right]\geq 0
\label{eq.entineq}
\end{split}
\end{align}
 where in the second line we use that $\D$ is diagonal, $\mathbf{v}(\x)=\J(\x)/p(\x)$, and $\Jr(\xr)=p(\xr) \int d\xl\ \mathbf{v}_{\rm r}(\x) p(\xl|\xr)= p(\xr) \mean{\mathbf{v}_{\rm r}(\x)|\xr}$, which follows from the derivation of Eq.~\eqref{eq:effectivefokker}.

\section{Derivation of Eq.~(\ref{eq:entropy2d})}
\noindent Here we derive the expression for the partial entropy production rate in terms of the cycling frequencies (see Eq.\eqref{eq:entropy2d}). It is convenient to substitute the current field $\mathbf{j}=\mathbf{\Omega}\x p(\x)$ in Eq.~\eqref{eq:entropyFokker}, which gives 
\begin{align}
\begin{split}
\Pi&=k_{\rm B}\int d\x (\Om \x)^T\D^{-1}(\Om \x)p(\x)= k_{\rm B} \int d\x \, x_i \Omega_{ij}^T (\D^{-1})_{jl} \Omega_{lm} x_m \\
&=k_{\rm B}\Omega_{ij}^T(\D^{-1})_{jl}\Omega_{lm} c_{mi}=k_{\rm B} \Tr{(\Om^T \D^{-1}\Om \C)}.
\end{split}
\end{align}
Since the entropy production is invariant under coordinate transformations,  we can use a more suitable coordinate system. In particular, we  choose  a set of coordinates such that $\C=\mathbb{1}$. In this set of coordinates, the entries of the matrix $\Omega _{ij}$ correspond to the cycling frequencies in the coordinates space of the $i^{th}$ and $j^{th}$ coordinates~\cite{Weiss2003}. Thus, in the 2D case $\Omr$ is given by
\begin{equation}
\label{eq:omr}
\Omr =\begin{pmatrix}
0 &\omega \\
-\omega & 0
\end{pmatrix}
\end{equation}
Furthermore, in this coordinate system $\Cr$ and $\Omr$ commute, yielding
\begin{equation}
\Pi_{\rm r}^{\rm 2D}=k_B\omega ^2\Tr{(\Cr \Dr^{-1})}
\end{equation}
Note, this expression is invariant under coordinate transformations.

\end{appendices}
\end{document}